\documentclass[twocolumn,pra,aps]{revtex4}
\usepackage{bm}
\usepackage{epsfig}

\begin{document}

\title{Model of molecular bonding based on the Bohr-Sommerfeld picture of
atoms}
\author{Anatoly A. Svidzinsky$^{a,b}$, Siu A. Chin$^a$ and Marlan O. Scully$%
^{a,b}$}
\affiliation{$^a$Institute for Quantum Studies and Department
 of Physics, Texas A\&M
University, TX 77843-4242 \\
$^b$Princeton Institute for the Science and Technology of Materials and
Dept. of Mechanical and Aerospace Engineering, Princeton University, NJ
08544 }
\date{\today }

\begin{abstract}
We develop a model of molecular binding based on the Bohr-Sommerfeld
description of atoms together with a constraint taken from conventional
quantum mechanics. The model can describe the binding energy curves of H$_2$%
, H$_3$ and other molecules with striking accuracy. Our approach treats
electrons as point particles with positions determined by extrema of an
algebraic energy function. Our constrained model provides a physically
appealing, accurate description of multi-electron chemical bonds.
\end{abstract}

\keywords{molecules, chemical bond, Bohr model}
\maketitle

Quantum chemistry has achieved excellent agreement between theory and
experiment by use of computational power to provide an adequate description
electron-electron interactions \cite{Scha84}. The conventional treatment of
molecular structure are based on solving the many-particle Schr\"odinger
equation with varying degree of sophistication, ranging from Diffusion Monte
Carlo methods, coupled cluster expansion, configuration interactions, to
density functional theory. All are intensely numerical, limited to rather
small systems and at the expense of providing a simple physical picture of
the chemical bond. Despite the successes of modern computational chemistry,
there remains a need for understanding electron correlations in some
relatively simple way so that we may describe ground and excited states of
large systems with reasonable accuracy.

Our goal here is to advance an intuitively appealing model of molecular
bonding capable of producing binding energy curves at chemical accuracy of a
few milli-Hartree. Our approach is based on the recently resurrected Bohr's
1913 model for molecules \cite{Bohr1}, which is derivable from an infinite
dimensional reduction of the Schr\"odinger equation \cite{Svid05a}. The
resulting electron configurations are reminiscent of the Lewis electron-dot
structure introduced in 1916 \cite{Lewi16}. The surprising feature of our
work is that all molecular binding energy curves studied below can be
accounted for by mostly electrostatic interaction of ``well-placed"
electrons, as if all the complicated kinetic and overlapping integrals have
been approximated by their mean-values via well-chosen electron positions.
Such an approach can potentially describe the structural elements of large
molecules beyond the current capability of ab initial methods.

\begin{figure}[tbp]
\bigskip
\centerline{\epsfxsize=0.3\textwidth\epsfysize=0.18\textwidth
\epsfbox{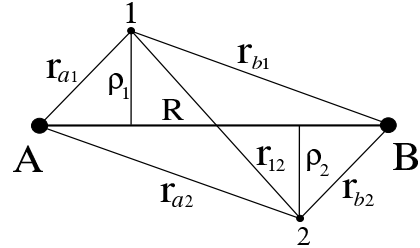}}
\caption{Electronic distances in H$_2$ molecule. The nuclei A and B are
fixed a distance $R$ apart. }
\label{coord}
\end{figure}

We will first derive our model for H$_2$, followed by applications to more
complex molecules. Fig. \ref{coord} displays various electron distances in H$%
_2$.
Distances and energies
are in units of the Bohr radius $a_0$ ($=\hbar ^2/me^2$) and Hartree ($%
=e^2/a_0)$ respectively. The original molecular Bohr model \cite{Bohr1}
quantize the electron's angular momentum
about the molecular axis resulting in
the ground state energy function \cite{Harc82,Svid05b},
\begin{equation}  \label{b1}
E=\frac 12\left( \frac 1{\rho _1^2}+\frac 1{\rho _2^2}\right) +V,
\end{equation}
where the first term is the Bohr kinetic energy and $V$ is the Coulomb
potential energy given in terms of electron distances defined in Fig. \ref%
{coord}:
\begin{equation}  \label{b2}
V=-\frac 1{r_{a1}}-\frac 1{r_{b1}}-\frac 1{r_{a2}}-\frac 1{r_{b2}}+\frac
1{r_{12}}+\frac 1R,
\end{equation}
$R$ is the internuclear separation. In our model, electron configurations of
a physical state correspond to extrema of an energy function, such as Eq. (%
\ref{b1}) \cite{Svid05a,Svid05b}.

\begin{figure}[h]
\bigskip
\centerline{\epsfxsize=0.45\textwidth\epsfysize=0.4\textwidth
\epsfbox{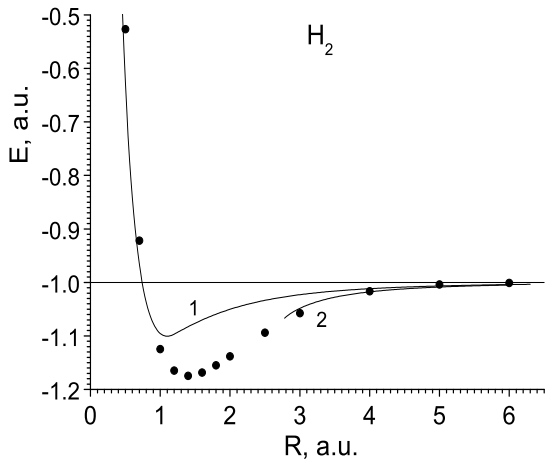}}
\caption{ Potential energy curve of the ground state of the H$_2$ molecule
obtained from the Bohr model with molecular axis quantization (curve 1) and
quantization relative to the nearest nucleus (curve 2). Solid circles are
the ``exact" energies \protect\cite{dot}. }
\label{coll}
\end{figure}

\begin{figure}[tbp]
\bigskip
\centerline{\epsfxsize=0.23\textwidth\epsfysize=0.28\textwidth
\epsfbox{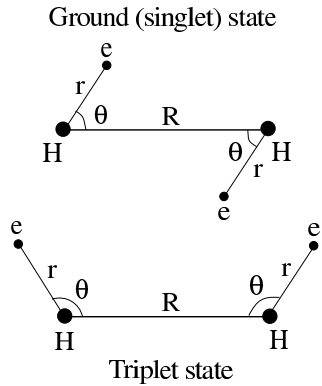}}
\caption{ Electron configuration for the ground (singlet) and triplet states
of H$_2$ molecule. }
\label{H2st}
\end{figure}

In Fig. \ref{coll} (curve 1) we plot the ground state $E(R)$ of Eq. (\ref{b1}%
), together with \textquotedblleft exact\textquotedblright\ quantum
mechanical results \cite{dot} (solid circles). The original Bohr model
yields a fairly accurate H$_{2}$ ground state energy $E(R)$ at small $R$,
but becomes increasely inaccurate at larger internuclear separations. This
can be remedied by the following observation. At large $R$ each electron in H%
$_{2}$ feels only the nearest nuclear charge, resulting in two weakly
interacting neutral H atoms. Therefore, at large $R$ each electron's angular
momentum ought to be quantized relative to the nearest nucleus, rather than
to the molecular axis. This asymptotic consideration yields the following H$%
_{2}$ energy
\begin{equation}
E=\frac{1}{2}\left( \frac{1}{r_{a1}^{2}}+\frac{1}{r_{b2}^{2}}\right) +V.
\label{b3}
\end{equation}%
For $R>2.77$ this energy function has a local minimum for the top
configuration of Fig. \ref{H2st}. We plot the corresponding $E(R)$ in Fig. %
\ref{coll} (curve 2). However, at $R<2.77$ the local minimum of the energy
function (\ref{b3}) disappears and each electron can collapse onto the
\textit{other} nucleus, \textit{i.e.}, $r_{b1}$ and/or $r_{a2}$ can vanish.
As one can see from Fig. \ref{coll}, the energy function (\ref{b3}), which
is a natural generalization of Bohr's hydrogen atom to the molecular case,
is in good quantitative agreement with the \textquotedblleft
exact\textquotedblright\ energy over the range of
$R>2.77$ where the local
minimum exists. This encourages us to seek a way of extending the
applicability of Eq. (\ref{b3}) to the entire range of $R$.

In the above naive generalization of Bohr's atom to the molecular case, each
electron can collapse onto the other nucleus because there is no
corresponding Bohr kinetic energy term about that nucleus to prevent the
collapse. By incorporating further insights from quantum mechanics, we can
remove this instability by a simple constraint. Quantum mechanically, the
two electrons are described by a wave function $\Psi (\mathbf{r}_{1},\mathbf{%
r}_{2})$. Electron 1 is a charge cloud with a most probable radius $r$. Let
\begin{equation}
\Phi (r,R)\equiv \left\langle \Psi \left\vert -\frac{1}{r_{\text{b1}}}%
\right\vert \Psi \right\rangle   \label{qmpot}
\end{equation}%
be the quantum mechanical potential between the electron cloud centered at
nucleus A and the nuclear charge of B, or vice versa. In the Bohr picture we
treat the electron as a point particle on a sphere of radius $r$ centered
about nucleus A. A subset of the spherical surface, a \textquotedblleft
circle\textquotedblright\ of positions $r$ satisfying
\begin{equation}
-\frac{1}{r_{\text{b1}}}=\Phi (r,R)  \label{a1}
\end{equation}%
will give the correct quantum mechanical interaction energy with nucleus B.
Thus if we impose the above as a constraint, and choose the electron
location only from this subset of the positions, then $r_{b1}$ can never be
zero, because the expectation value in Eq. (\ref{qmpot}) is finite.

One can derive the effective potential $\Phi (r,R)$ from any simple
two-electron wave functions, such as the Heitler-London (HL) \cite{Heit27}
or the Hund-Mulliken (HM) \cite{HM}
wave function $\Psi $. The HL wave function is
\begin{equation}  \label{c1}
\Psi =a(1)b(2)\pm b(1)a(2),
\end{equation}
where
``+/-" corresponds to singlet/triplet state and
\[
a(i)=\sqrt{\frac{\alpha ^3}\pi }\exp (-\alpha r_{ai}),\quad b(i)=\sqrt{\frac{%
\alpha ^3}\pi }\exp (-\alpha r_{bi}),
\]
for $i=1-2$ are variational wave functions with parameter $\alpha$. If we
take $a(1)$ as a variational wave function for an isolated hydrogen atom,
then the variational energy is $E=\alpha ^2/2-\alpha$. This
reduces to the Bohr model energy function for the hydrogen atom if we
identify $\alpha =1/r$, where $r$ is the radial distance from the nucleus.
We will also use $r$ to denote the radial distance of an electron from its
nearest nucleus.

For the HL wave function Eq. (\ref{qmpot}) yields the well known Coulomb and
exchange integral,
\begin{equation}  \label{c3}
\Phi =-\frac 1{1\pm S^2}\left\{ \int a^2(1)\frac 1{r_{\text{b1}}}d\mathbf{r}%
_1\pm 2S\int a(1)b(1)\frac 1{r_{\text{b1}}}d\mathbf{r}_1\right\} ,
\end{equation}
with overlap $S=\int a(1)b(1)d\mathbf{r}_1.$ More explicitly, the singlet
and triplet potentials are respectively
\begin{equation}  \label{a2}
\Phi _{\text{s}}(r,R)=-\frac 1{1+S^2(r,R)} \left[ f(r,R) +S(r,R) g(r,R) %
\right] ,
\end{equation}
\begin{equation}  \label{a5}
\Phi _{\text{t}}(r,R)=-\frac 1{1-S^2(r,R)} \left[f(r,R) -S(r,R) g(r,R) %
\right] ,
\end{equation}
with $f(r,R)=1/R -\exp (-2R/r)\left(1/r+1/R\right)$, $g(r,R)=\exp
(-R/r)(1+R/r)/r$ and \hfill
\begin{equation}  \label{a3}
S(r,R)=\exp (-R/r)\left( 1+\frac Rr+\frac{R^2}{3r^2}\right) .
\end{equation}
The singlet state Hund-Mulliken wave function $\Psi =[a(1)+b(1)][a(2)+b(2)]$
yields the HM singlet effective potential:%
\begin{equation}  \label{hmsing}
\Phi _{\text{s}}(r,R)=-\frac 1{1+S(r,R)} \left[f(r,R)+g(r,R) \right] .
\end{equation}
For the triplet state, the HL and HM wave functions are the same, yielding
identical triplet potentials. Note that by introducing singlet and triplet
potentials, we have augmented the original atomic Bohr model with elements
of Pauli's exclusion principle. The latter is essential for any successful
description of atomic and molecular systems.

When we apply the constrained Bohr model to H$_{2}$, the resulting energy
function is
\begin{equation}
E(r,R)=\frac{1}{r^{2}}-\frac{2}{r}+2\Phi (r,R)+\frac{1}{r_{\text{12}}}+\frac{%
1}{R}.  \label{c5}
\end{equation}%
The energy function has an extremum when $r_{a1}=r_{b2}=r$ and $r_{a2}=r_{b1}
$. The resulting electron configurations corresponding to the ground and
triplet states are as shown in Fig. \ref{H2st}, where, for singlet ground
state%
\[
r_{\text{12}}=\sqrt{2r^{2}-R^{2}+\frac{2}{\Phi ^{2}(r,R)}},
\]%
and for the triplet excited state,
\[
r_{\text{12}}=\frac{1}{R\Phi ^{2}(r,R)}-\frac{r^{2}}{R}.
\]%
These are just geometric distances between the two electrons expressed in
terms of $R$, $r$ and $r_{b1}$. The angle $\theta $ is determined by the
relation $1/\Phi ^{2}=R^{2}+r^{2}-2rR\cos \theta $.

\begin{figure}[tbp]
\bigskip
\centerline{\epsfxsize=0.5\textwidth\epsfysize=0.45\textwidth
\epsfbox{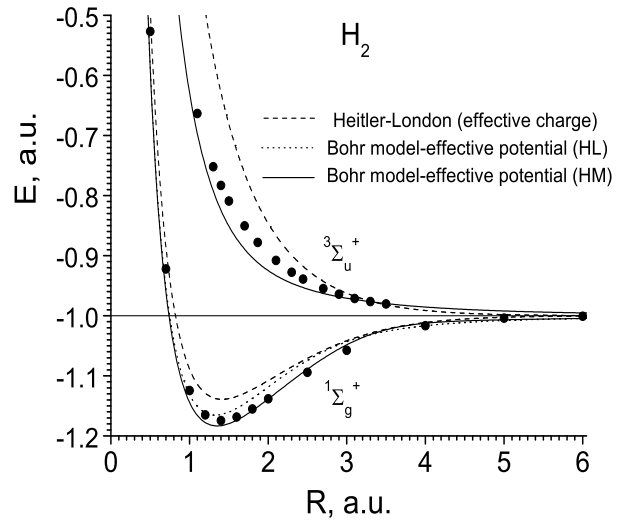}}
\caption{ Potential energy curves of the ground $^1\Sigma _g^{+}$ and first
triplet state $^3\Sigma _u^{+}$ of the H$_2$ molecule. Solid lines are
obtained from the constrained Bohr model with HM effective potential, while
the small dot line is derived with HL potential. Dashed curves are from HL
effective charge variational treatment. }
\label{h2eff}
\end{figure}

The binding energy curves $E(R)$ for both singlet and triplet states are
shown in Fig. \ref{h2eff}. There are no fitting parameters in our
calculations. The solid and dotted lines are results from using the HM and
HL potential respectively. Solid circles are \textquotedblleft
exact\textquotedblright\ results \cite{dot}. The constrained Bohr model
gives a surprisingly accurate $E(R)$ at all $R$, yielding a ground state
binding energy of $E_{B}=4.50$ eV for the HL potential and $E_{B}=4.99$ eV
for the HM potential \cite{note}. The \textquotedblleft
exact\textquotedblright\ result is $E_{B}=4.745$ eV \cite{Scha84}. The
Heitler-London-Wang effective charge calculation
(dashed curves) gives substantially worse accuracy
with $E_{B}=3.78$ eV \cite{Heit27,Wang28}.
Only more elaborate variational calculations with configuration interaction
can produce energies comparable to our constrained Bohr model results.

Generalizing the constraint Eq. (\ref{a1}) to a system of several hydrogen
atoms is straightforward. Let's consider electron $1$ belonging to its
nearest nucleus $1$ and denote the distances from
electron $1$ to nuclei $i$ as $r_i$ ($i=1$, $2$, $\ldots $ ). Then the
constraint equation reads
\begin{equation}  \label{a6}
-\sum_{i>1}\frac 1{r_i}=\sum_{i>1}\Phi _i(r_1,R_i),
\end{equation}
where $R_i$ is the separation between nucleus $1$ and nucleus $i$. Mutual
spin orientation of electrons $1$ and electron $i$ (belonging to nucleus $i$%
) determines a singlet or triplet $\Phi _i$ in Eq. (\ref{a6}). In this way,
we have incorporated elements of Pauli's exclusion principle into our model.

\begin{figure}[!]
\includegraphics[angle=0,width=8.5cm]{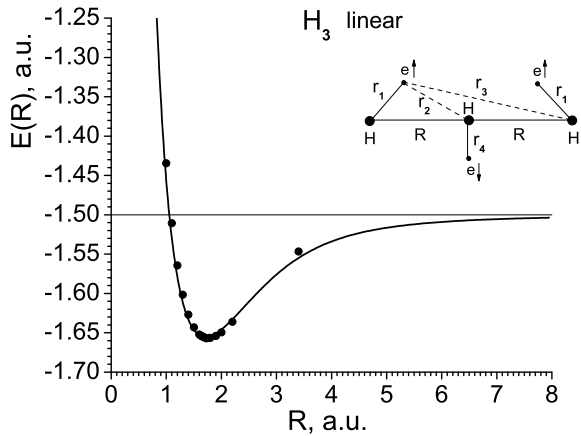}
\caption{Electron configuration and the ground state $E(R)$ of the linear H$%
_3$ molecule obtained from the constrained Bohr model (solid curve) and
``exact" numerical solution of the Schr\"odinger equation (solild circles). }
\label{H3l}
\end{figure}

\begin{figure}[h]
\bigskip
\centerline{\epsfxsize=0.5\textwidth\epsfysize=0.45\textwidth
\epsfbox{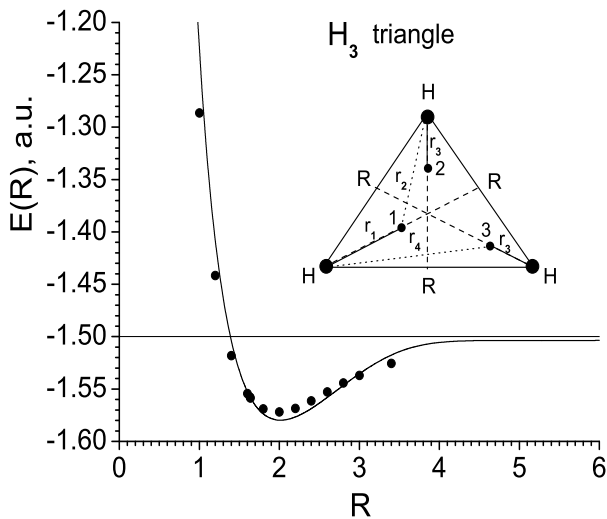}}
\caption{Electron configuration and the ground state $E(R)$ of the
triangular H$_3$ molecule. Solid curve is the result of the constrained Bohr
model while solid circles are the ``exact" numerical answer. }
\label{H3t}
\end{figure}

When applying our model to the triatomic H$_3$ molecule, we consider linear
and triangular configurations as shown in the insert of Figs. \ref{H3l} and %
\ref{H3t}. The spacing between the nearest nuclei is assumed to be the same,
equal to $R$. Due to symmetry, the central electron in the linear H$_3$
molecule must be at equal distances from the two neighboring nuclei. For
this electron, since its position is fixed, there is no collapse and
therefore no need for any constraint. We only need to constrain the two
outermost electrons (see insert of Fig. \ref{H3l}) via Eq. (\ref{a6}) in the
form
\begin{equation}  \label{a7}
-\frac 1{r_2}-\frac 1{r_3}=\Phi _s(r_1,R)+\Phi _t(r_1,2R),
\end{equation}
where $r_i$ are defined in the insert of Fig. \ref{H3l}. For the linear H$_3$
ground state, adjacent electrons in the molecule have opposite spins,
requiring the singlet potential $\Phi _s(r_1,R)$. (We use the HM singlet
potential given by Eq. (\ref{hmsing})). In this case, the spins of the two
outermost electrons must be parallel requiring the triplet potential $\Phi
_t(r_1,2R)$. Minimizing the resulting energy function
\begin{equation}  \label{a8}
E=\frac 1{r_1^2}+\frac 1{2r_4^2}+V
\end{equation}
yields the solid line potential energy curve of Fig. \ref{H3l}, which
essentially goes through the ``exact'' solid circle results.

\begin{figure}[h]
\bigskip
\centerline{\epsfxsize=0.45\textwidth\epsfysize=0.4\textwidth
\epsfbox{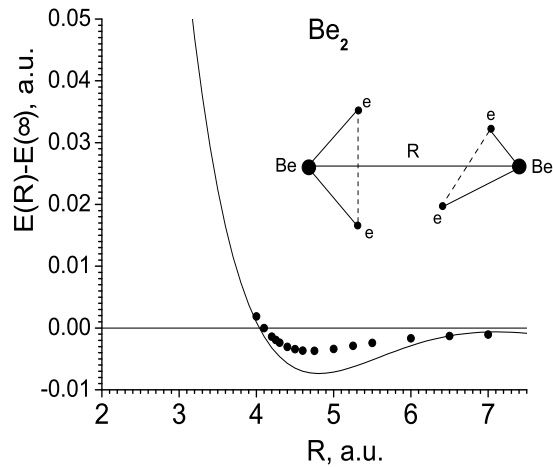}}
\caption{Configuration of outer electrons and the ground state $E(R)$ of the
Be$_2$ molecule obtained in the constrained Bohr model (solid curve) and the
``exact" result (solid circles). The Bohr model provides accuracy of $0.003$
Hartree. }
\label{Be2}
\end{figure}

Insert of Fig. \ref{H3t} shows the electrons' positions for the equilateral
triangle nuclei configuration. We assume that electron 1 has spin opposite
to those of electrons 2 and 3. Symmetry dictates that electron 1 lies above,
while electrons 2 and 3 lie below the nuclear plane along the bisector of
the equilateral triangle.
For electron 1 the constraint Eq. (\ref{a6}) reads
\begin{equation}
-\frac{1}{r_{2}}=\Phi _{s}(r_{1},R),  \label{a9}
\end{equation}%
while for electrons 2 and 3 we have
\begin{equation}
-\frac{2}{r_{4}}=\Phi _{s}(r_{3},R)+\Phi _{t}(r_{3},R).  \label{a10}
\end{equation}%
Minimization of the
energy function
\begin{equation}
E=\frac{1}{2r_{1}^{2}}+\frac{1}{r_{3}^{2}}+V  \label{a11}
\end{equation}%
with the constraints (\ref{a9}) and (\ref{a10}) results in the solid line
potential energy curve as shown in Fig. \ref{H3t}. Again for $\Phi _{s}$ we
take the HM effective potential given by Eq. (\ref{hmsing}). As in the case
of the linear H$_{3}$ molecule, the constrained Bohr model yielded very
accurate $E(R)$ over the entire range of internuclear separation. The
constrained Bohr model also gives good results for other molecules, e.g., Be$%
_{2}$ as shown in Fig. \ref{Be2} and H$_{4}$.

In the vicinity of the energy minimum
the constrained Bohr model for Be$_{2}$
provides accuracy of a few milli Hartree with no fitting parameters.
However, since Be$_{2}$ is very weakly bound the binding energy is
off by about 50\%.
Nevetheless, the bond length remains quite accurate.

In summary, we have shown that the atomic Bohr model, when supplemented by
potential constraints obtained from quantum mechanics incorporating Pauli's
exclusion principle, gives a remarkably accurate description of diatomic and
triatomic molecules. This constrained Bohr model provides a physically
appealing geometric picture of how multi-electron bonds are formed and holds
promise for future applications to complex many-electron systems. Possible
application of the model includes the simulation of biological molecules, where
there is no ab initio way of doing the calculations with any other methods.
The model can also be applied to the calculation of
potential curves of molecules in a super strong magnetic field
on surfaces of white dwarf and neutron stars.

This work was supported by the Robert A. Welch Foundation Grant A-1261,
Office of Naval Research (Award No. N00014-03-1-0385)
and by a National Science Foundation grant (to SAC) DMS-0310580.


\begin{thebibliography}{99}
\bibitem{dot} Throughout this paper ``exact'' energies (solid circles) are
taken from a combination of experimental and theoretical sources. The
theoretical analyses are typically many term (10-100) configuration
interaction computer calculations, see, for example, Ref. \cite{Scha84}.
\bibitem{note} Since the Bohr model description is not fully quantum
mechanical it disobeys the variational principle which can yield energy
over-binding.
\end{thebibliography}
\end{document}